\documentclass[aps,prc,preprint,tightenlines,groupedaddress,nofootinbib,
showpacs,preprintnumbers,amsmath,amssymb,superscriptaddress]{revtex4}
\usepackage{graphicx}
\usepackage{dcolumn}
\usepackage{mathrsfs}
\usepackage{bm}

\begin{document}

\title{The Nuclear Physics of Neutron Stars}
\author{J. Piekarewicz}
\affiliation{Department of Physics, Florida State
             University, Tallahassee, FL 32306}
\date{\today}

\begin{abstract}
A remarkable fact about spherically-symmetric neutron stars in 
hydrostatic equilibrium --- the so-called Schwarzschild stars 
--- is that the only physics that they are sensitive to is the 
equation of state of neutron-rich matter. As such, neutron stars 
provide a myriad of observables that may be used to constrain 
poorly known aspects of the nuclear interaction under extreme 
conditions of density. After discussing many of the fascinating 
phases encountered in neutron stars, I will address how powerful 
theoretical, experimental, and observational constraints may be 
used to place stringent limits on the equation of state of
neutron-rich matter.
\end{abstract}
\maketitle

\section{Introduction}
\label{Introduction}

A neutron star is a gold mine for the study of the phase diagram of
cold baryonic matter. While the most common perception of a neutron
star is that of a uniform assembly of neutrons packed to densities
that may exceed that of normal nuclei by up to an order of magnitude, 
the reality is far different and significantly more interesting. 
Indeed, the mere fact that hydrostatic equilibrium must be maintained 
throughout the neutron star, demands a {\sl negative} pressure gradient 
at each point in the star; otherwise the star would collapse under its 
own weight. This model-independent fact yields nuclear densities --- 
at least for most realistic equations of states --- that span over 11 
orders of magnitude; from $10^{4}$ to $10^{15}~{\rm g/cm^{3}}$. Recall
that in this units nuclear-matter saturation density equals
$\rho_{0}\!=\!2.48\times 10^{14}{\rm g/cm^{3}}$.
What novels phases of baryonic matter emerge under these conditions is
both fascinating and unknown. Moreover, most of the exotic phases
predicted to exist in neutron stars can not be realized under normal
laboratory conditions. Whereas most of these phases have a fleeting
existence here on Earth, they become stable in neutron stars as a
consequence of the presence of enormous gravitational fields.

To establish the fundamental role played by the equation of state on
the structure of spherically-symmetric neutron stars in hydrostatic
equilibrium, we start with the Tolman-Oppenheimer-Volkoff (TOV)
equations --- the extension of Newton's laws to the domain of general 
relativity. The TOV equations may be expressed as a coupled set of 
first-order differential equations of the following form:
\begin{subequations}
 \begin{align}
   & \frac{dP}{dr}=-G\,\frac{{\cal E}(r)M(r)}{r^{2}}
         \left[1+\frac{P(r)}{{\cal E}(r)}\right]
         \left[1+\frac{4\pi r^{3}P(r)}{M(r)}\right]
         \left[1-\frac{2GM(r)}{r}\right]^{-1} \;,
         \label{TOVa}\\
   & \frac{dM}{dr}=4\pi r^{2}{\cal E}(r)\;,
         \label{TOVb}
 \end{align}
 \label{TOV}
\end{subequations}
where $G$ is Newton's gravitational constant, while $P(r)$, 
${\cal E}(r)$, and $M(r)$ represent the pressure, energy density, 
and enclosed-mass profiles of the star, respectively. Note that the
last three terms (enclosed in square brackets) in Eq.~(\ref{TOVa}) 
have a general-relativistic origin. Remarkably, the only input
that neutron stars are sensitive to is the equation of state of 
neutron-rich matter. Indeed, changes in pressure and enclosed mass
as a function of radius (left-hand side of the equations) depend
not only on the values of these quantities at $r$, but also on the
``unknown'' energy density ${\cal E}(r)$ of the system. Thus, no 
solution of the TOV equations is possible until an equation of state 
({\sl i.e.,} a $P$ {\sl vs} ${\cal E}$ relation) is supplied.

In this manuscript we discuss the various fascinating phases of
baryonic matter that are predicted to exist in neutron stars, but
inaccessible under normal laboratory conditions.  After briefly
discussion the theoretical models used in this contribution, we focus
on recent theoretical, experimental, and observational constrains that
place stringent limits on the equation of state of neutron-rich
matter.

\section{Anatomy of a Neutron Star}
\label{Anatomy}

\begin{figure}[h]
  \includegraphics[height=2.7in]{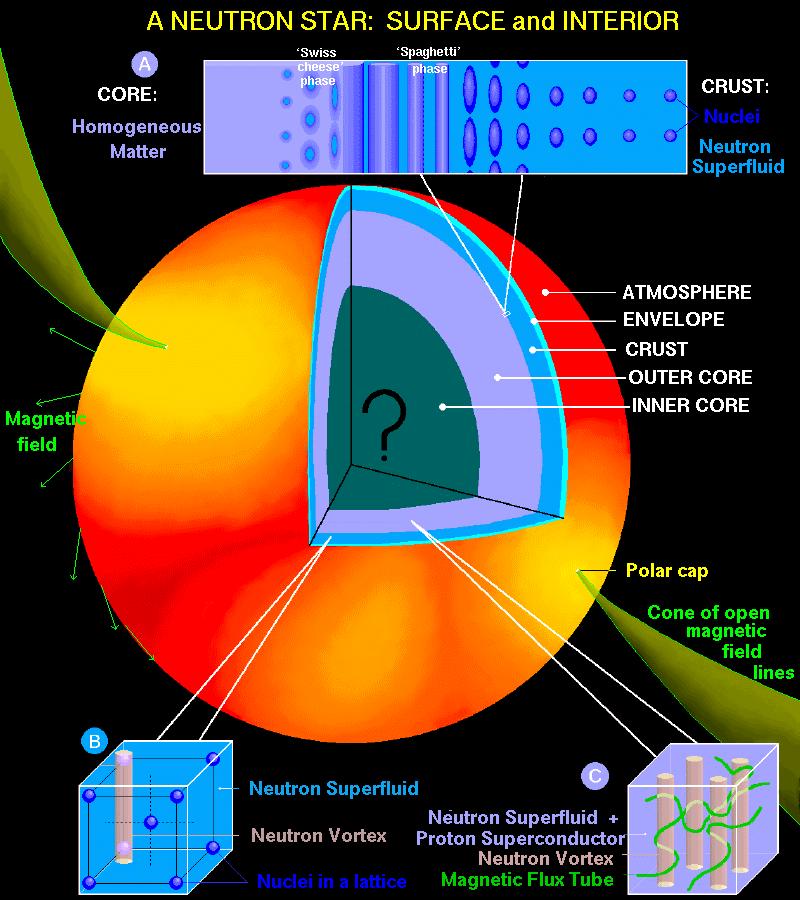}
 \vspace{-0.2cm}
 \caption{A rendition of the structure and phases 
          of a neutron star (courtesy of Dany Page).}
 \label{Fig1}
\end{figure}

Neutron stars contain a non-uniform crust above a uniform liquid
mantle. See Fig.~\ref{Fig1} for what is believed to be an accurate
rendition of the structure of a neutron star.

\subsection{The Outer Crust}
\label{OuterCrust}

The outer crust is understood as the region of the star spanning about
7 orders of magnitude in density; from about $10^{4}{\rm g/cm^{3}}$ to
$4\times 10^{11}{\rm g/cm^{3}}$~\cite{Baym:1971pw}. At these densities, 
the electrons --- which are an essential
component of the star in order to maintain charge neutrality --- have
been pressure ionized and move freely throughout the crust. Moreover,
at these ``low'' densities, ${}^{56}$Fe nuclei arrange themselves in a
crystalline lattice in order to minimize their overall Coulomb
repulsion. This is the structure of the outermost layer of the crust. 
However, as the density increases (and one moves away from the
surface of the star) ${}^{56}$Fe is no longer the most energetically
favorable nucleus.  This is because the electronic contribution to the
energy increases faster with density than the nuclear contribution. As
a result, it becomes energetically advantageous for the energetic
electrons to capture on the protons and for the excess energy to be
carried away by neutrinos. The resulting nuclear lattice is now made
of nuclei having a neutron excess larger than that of ${}^{56}$Fe. As
the density continues to increase, the nuclear system evolves into a
Coulomb lattice of progressively more neutron-rich nuclei until a
{\sl ``critical''} density of about $4\times 10^{11}{\rm g/cm^{3}}$ 
is reached. At this point the nuclei are unable to hold any more 
neutrons; the neutron drip line has been reached.

\subsection{The Inner Crust}
\label{Inner Crust}

\begin{figure}[h]
  \includegraphics[height=2.5in]{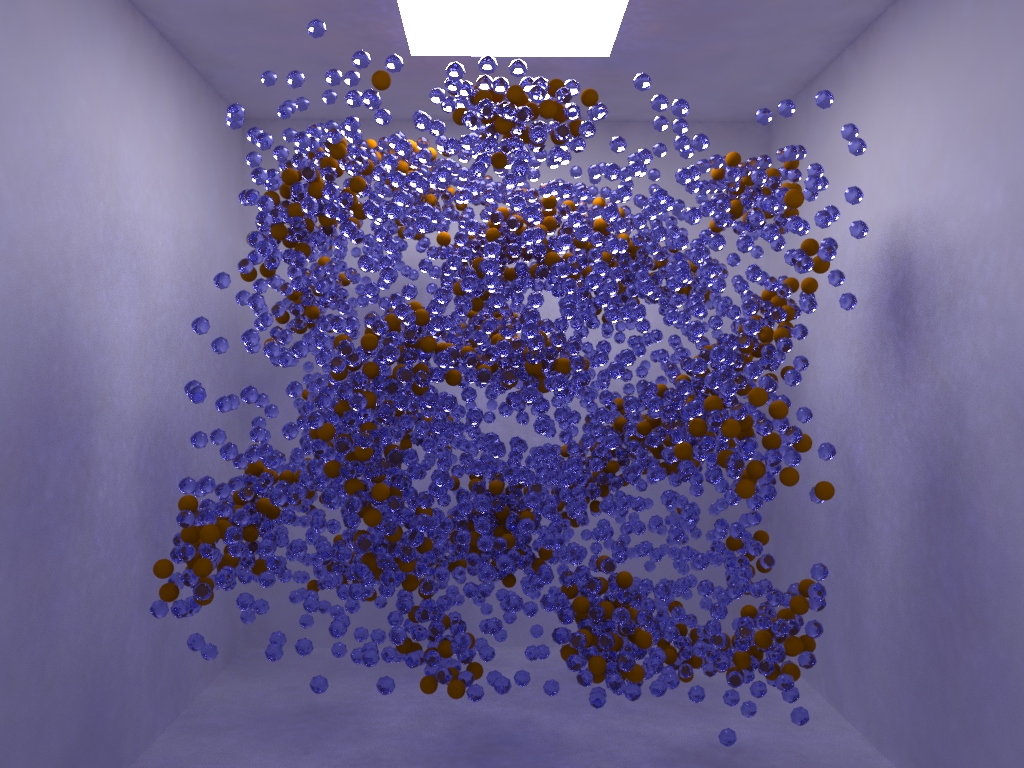}
 \vspace{-0.2cm}
 \caption{(color online) A snapshot of a Monte Carlo simulation 
          for a configuration of 4,000 nucleons at a baryon 
          density of  $n\!=\!0.025~{\rm fm}^{-3}$ (a sixth of 
          normal nuclear matter saturation density), a proton 
          fraction of $Y_{p}\!=\!Z/A\!=\!0.2$, and a temperature 
          of $T\!=\!1$~MeV.} 
 \label{Fig2}
\end{figure}

The inner crust of the neutron star comprises the region from
neutron-drip density up to the density at which uniformity in the
system is restored (approximately $1/3$ to $1/2$ of normal nuclear
matter saturation density). At these densities the system exhibits
rich and complex structures that emerge from a dynamical competition
between short-range nuclear attraction and long-range Coulomb
repulsion. At the lower densities present in the {\sl outer core},
these length scales are well separated and the system organizes itself
into a crystalline lattice of neutron-rich nuclei. In contrast, at
a much higher density of the order of half of nuclear-matter 
saturation density, uniformity in the system is restored and the
system behaves as a uniform Fermi liquid.  Yet the transition region
from the highly-ordered crystal to the uniform liquid mantle is
complex and not well understood. Length scales that were well
separated in both the crystalline and uniform phases are now
comparable, giving rise to a universal phenomenon known as {\sl
``Coulomb frustration''}. It has been speculated that the transition
to the uniform phase {\sl must} go through a series of changes in the
dimensionality and topology of these complex structures, colloquially
known as {\sl ``nuclear pasta''}~\cite{Ravenhall:1983uh,
Hashimoto:1984}. In Fig.~\ref{Fig2} a snapshot obtained from
Monte-Carlo/Molecular-Dynamics simulations of a nuclear system at
densities relevant to the inner crust are
displayed~\cite{Horowitz:2004yf,Horowitz:2004pv}. The figure displays
how the system organizes itself into neutron-rich clusters ({\sl
i.e.,} ``nuclei'') of complex topologies that are surrounded by a 
vapor of (perhaps superfluid) neutrons. Such complex pasta structures 
may have a significant impact on various transport properties, such 
as neutrino and electron propagation.

\subsection{The Stellar Core}
\label{Core}

As the density continues to increase, the neutron-rich nuclei will
``melt'' and uniformity in the system will be restored.  At these
densities (of the order of $1/3$-to-$1/2$ of normal nuclear-matter
saturation density) the naive perception of a neutron star, namely, 
a uniform assembly of closely-packed neutrons will be realized, albeit
for the presence of a small percentage (of the order of 10\%) of
protons and electrons required to maintain chemical equilibrium and
charge neutrality. Although the non-uniform crust displays fascinating
and intriguing dynamics, its structural impact on the star is rather
modest. Indeed, the crust accounts for about 10\% of the radius of the
neutron star and for only a few percent of its mass. Most of the mass
of the neutron star is contained in its uniform core.  Yet the highest
density attained in the core depends critically on the equation of
state of neutron-rich matter which at those high densities is poorly
constrained. The cleanest constraint on the equation of state at
high-density will emerge as we answer one of the central questions in
nuclear astrophysics: {\sl what is the maximum mass of a neutron
star?} Or equivalently, {\sl what is the minimum mass of a black
hole?} Note that if the equation of state is ``soft'', very high
densities may be reached in the stellar core. At such high densities 
new states of matter may develop as the quarks within the hadrons 
become  deconfined. Such an exciting possibility will not be 
considered further in this manuscript.

\section{Constraints on the Equation of State}
\label{Constraints}

Before addressing the role that recent observables play in constraining 
various theoretical description of the equation of state, we introduce
the relativistic mean-field models that are used to compute these
observables. 

Relativistic mean-field descriptions of the ground-state properties of
medium to heavy nuclei have enjoyed enormous success. These highly
economical descriptions encode a great amount of physics in a handful
of model parameters that are calibrated to a few ground-state
properties of a representative set of medium to heavy nuclei. An
example of such a successful paradigm is the relativistic NL3
parameter set of Lalazissis, Ring, and
collaborators~\cite{Lalazissis:1996rd,Lalazissis:1999}.

The Lagrangian density employed in this work is rooted on the seminal
work of Walecka, Serot, and their many collaborators (see
Refs.~\cite{Walecka:1974qa,Serot:1984ey,Serot:1997xg} and references
therein). Since first published by Walecka more than three decades
ago~\cite{Walecka:1974qa}, several refinements have been implemented
to improve the quantitative standing of the model. In the present work
we employ an interacting Lagrangian density of the following
form~\cite{Mueller:1996pm,Horowitz:2000xj,Todd-Rutel:2005fa}:
\begin{widetext}
\begin{align}
{\mathscr L}_{\rm int} & =
 \bar\psi\left[g_{\rm s}\phi   \!-\!
         \left(g_{\rm v}V_\mu  \!+\!
    \frac{g_{\rho}}{2}\tau\cdot{\bf b}_{\mu}
                               \!+\!
    \frac{e}{2}(1\!+\!\tau_{3})A_{\mu}\right)\gamma^{\mu}
         \right]\psi \nonumber \\
                   & -
    \frac{\kappa}{3!} (g_{\rm s}\phi)^3 \!-\!
    \frac{\lambda}{4!}(g_{\rm s}\phi)^4 \!+\!
    \frac{\zeta}{4!}
    \Big(g_{\rm v}^2 V_{\mu}V^\mu\Big)^2 \!+\!
    \Lambda_{\rm v}
    \Big(g_{\rho}^{2}\,{\bf b}_{\mu}\cdot{\bf b}^{\mu}\Big)
    \Big(g_{\rm v}^2V_{\mu}V^\mu\Big) \;.
 \label{Lagrangian}
\end{align}
\end{widetext}
The original Lagrangian density of Walecka consisted of an isodoublet
nucleon field ($\psi$) together with neutral scalar ($\phi$) and
vector ($V^{\mu}$) fields coupled to the scalar density
($\bar\psi\psi$) and conserved nucleon current
($\bar\psi\gamma^{\mu}\psi$), respectively~\cite{Walecka:1974qa}. In
spite of its simplicity (indeed, the model contains only two
dimensionless coupling constants), symmetric nuclear matter saturates
even when the model was solved at the mean-field
level~\cite{Walecka:1974qa}. By adding additional contributions from a
single isovector meson ($b^{\mu}$) and the photon ($A^{\mu}$),
Horowitz and Serot~\cite{Horowitz:1981xw} obtained results for the
ground-state properties of finite nuclei that rivaled some of the most
sophisticated non-relativistic calculations of the time. However,
whereas the two dimensionless parameters in the original Walecka model
could be adjusted to reproduce the nuclear saturation point, the
incompressibility coefficient (now a prediction of the model) was too
large ($K\!\gtrsim\!500$~MeV) as compared with existing data on
breathing-mode energies~\cite{Youngblood:1977}. To overcome this
problem, Boguta and Bodmer introduced cubic ($\kappa$) and quartic
($\lambda$) scalar meson self-interactions that accounted for a
significant softening of the equation of state
($K\!=\!150\!\pm\!50$~MeV)~\cite{Boguta:1977xi}.  Two parameters of
the Lagrangian density of Eq.~(\ref{Lagrangian}) remain to be
discussed, namely, $\zeta$ and $\Lambda_{\rm v}$. Both of these
parameters are set to zero in the enormously successful NL3 model,
suggesting that the experimental data used in the calibration
procedure is insensitive to the physics encoded in these
parameters. Indeed, M\"uller and Serot found possible to build models
with different values of $\zeta$ that reproduce the same observed
properties at normal nuclear densities, but which yield maximum
neutron star masses that differ by almost one solar
mass~\cite{Mueller:1996pm}. This result indicates that observations of
massive neutron stars --- rather than laboratory experiments --- may
provide the only meaningful constraint on the high-density component
of the equation of state. Finally, the isoscalar-isovector coupling
constant $\Lambda_{\rm v}$ was added in Ref.~\cite{Horowitz:2000xj} to
modify the density dependence of the symmetry energy.  It was
found that models with different values of $\Lambda_{\rm v}$ reproduce
the same exact properties of symmetric nuclear matter, but yield
vastly different values for the neutron skin thickness of heavy nuclei
and for the radii of neutron stars~\cite{Horowitz:2001ya}.  The Parity
Radius Experiment (PREx) at the Jefferson Laboratory promises to
measure the skin thickness of $^{208}$Pb accurately and model
independently via parity-violating electron
scattering~\cite{Horowitz:1999fk, Michaels:2005}. PREx will provide a
unique experimental constraint on the density dependence of the
symmetry energy due its strong correlation to the neutron skin of
heavy nuclei~\cite{Brown:2000}.

\subsection{Theoretical Constraints}
\label{Theory}

One of the most stringent constraints on the equation of state of 
low density neutron-rich matter emerges from theoretical considerations, 
namely, from the universality of dilute Fermi gases with an ``infinite'' 
scattering length ($a$). In this limit the only energy scale in the 
problem is the Fermi energy ($\varepsilon_{\rm F}$), so the energy per 
particle is constrained to be that of the free Fermi gas up to a 
dimensionless {\sl universal constant} ($\xi$) that is independent 
of the details of the two-body interaction~\cite{Carlson:2003}. 
\begin{figure}[h]
  \includegraphics[width=4in]{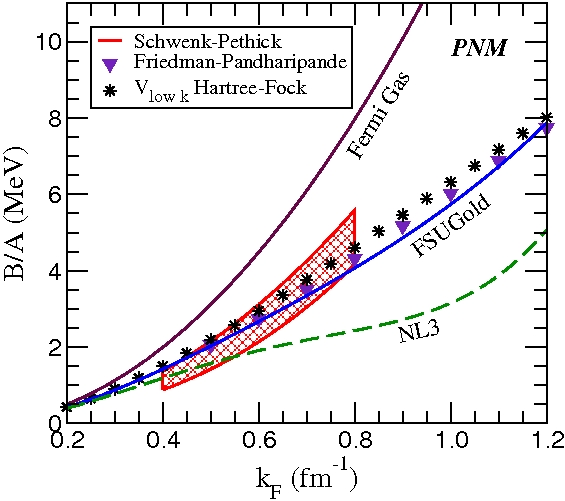}
 \vspace{-0.2cm}
 \caption{(color online) Equation of state of pure neutron matter
         as a function of the Fermi momentum. Predictions  are
         shown for the accurately calibrated
         NL3~\cite{Lalazissis:1996rd,Lalazissis:1999} (green
         line) and  FSUGold~\cite{Todd-Rutel:2005fa} (blue line)
         parameter sets. Shown also are various microscopic
         descriptions --- including a {\sl model-independent}
         result based on the physics of resonant Fermi gases by
         Schwenk and Pethick~\cite{Schwenk:2005ka} (red region).}
\label{Fig3}
\end{figure}
That is,
\begin{equation}
  \frac{E}{N} = \xi\frac{3}{5}\varepsilon_{\rm F} \;.
\end{equation}
To date, the best theoretical estimates place the value of the
universal constant around $\xi\!\approx\!0.4$~\cite{Baker:1999dg,
Heiselberg:2000bm,Carlson:2003,Nishida:2006br}.

Although the neutron-neutron scattering length is large indeed
($a_{\rm nn}\!=\!-18.5$~fm), pure neutron matter deviates from
unitarity due to a non-negligible value of the effective range 
of the neutron-neutron interaction ($r_{\rm e}\!=\!+2.7$~fm). Thus,
corrections to the low-density equation of state of pure neutron
matter must be computed for $k_{\rm F}\!\sim\!r_{\rm
e}^{-1}\simeq0.4~{\rm fm}^{-1}$. Such corrections have been recently
computed by Schwenk and Pethick~\cite{Schwenk:2005ka}, with their
results displayed as the red hatched region in Fig~\ref{Fig3}. Also
shown are the predictions of two microscopic models based on realistic
two-body interactions, one of them being the venerated equation of
state of Friedman and Pandharipande~\cite{Friedman:1981qw}. Finally,
the predictions of NL3 and FSUGold are also shown.  It is gratifying
that the softening of the symmetry energy of FSUGold --- caused by
incorporating constraints from breathing-mode
energies~\cite{Todd-Rutel:2005fa} --- appears consistent with the
physics of resonant Fermi gases.  Such a powerful universal constraint
should be routinely and explicitly incorporated into future
determinations of density functionals. Indeed, such a constrain
appears to rule out many of the models displayed in Fig.~2 of
Ref.~\cite{Brown:2000}.

\subsection{Experimental Constraints}
\label{Experiment}

\begin{figure}[h]
  \includegraphics[height=3.5in,width=4in]{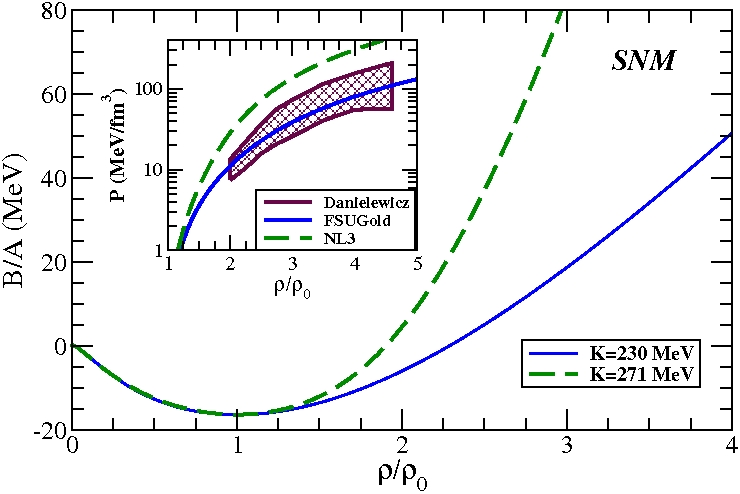}
 \vspace{-0.2cm}
\caption{(color online) Binding energy per nucleon as a function
         of baryon density (expressed in units of the saturation
         density $\rho_{0}\!=\!0.148~{\rm fm}^{-3}$) for symmetric
         nuclear matter. Theoretical predictions are shown for the
         NL3~\cite{Lalazissis:1996rd,Lalazissis:1999} (green line)
         and FSUGold~\cite{Todd-Rutel:2005fa} (blue line) models.
         Shown in the inset is a comparison between the equation
         of state extracted from energetic nuclear
         collisions~\cite{Danielewicz:2002pu} and the predictions
         of these two models.}
\label{Fig4}
\end{figure}

Energetic nuclear collisions may be used to constrain the high-density
behavior of nucleonic matter. To illustrate this point we display in
Fig.~\ref{Fig4} the binding energy per nucleon of {\sl symmetric}
nuclear matter as a function of the baryon density as predicted by
both the NL3 and FSUGold models. Note that both models reproduce the
equilibrium properties of symmetric nuclear matter and display the
same {\sl quantitative} behavior at densities below the saturation
point.  Yet their high-density predictions are significantly
different.  This emerges from a combination of two factors. First,
FSUGold predicts an incompressibility coefficient $K$ considerably
lower than NL3, namely, $230$~MeV {\sl vs} $271$~MeV. Second, and more
importantly, FSUGold includes an omega-meson self-energy coupling 
[labeled by $\zeta$ in Eq.~(\ref{Lagrangian})] that is responsible for a 
significant softening at high density. We now compare the predictions
of these two models against results obtained from energetic nuclear
collisions that can compress baryonic matter to densities as high as
those predicted to exist in the core of neutron stars. The inset in
Fig.~\ref{Fig4} provides us with such a comparison. By analyzing the
manner in which matter flows after the collision of two energetic gold
nuclei, the equation of state of {\sl symmetric} nuclear matter was
extracted up to densities of 4-to-5 times saturation
density~\cite{Danielewicz:2002pu}. Figure~\ref{Fig4} seems to rule out
overly stiff equations of state (such as NL3). And while it is
gratifying that FSUGold is consistent with this analysis, one
must stress that the connection between energetic nuclear collisions 
and the equation of state of cold nuclear matter is model dependent.

\subsection{Observational Constraints}
\label{Observation}

\begin{figure}[h]
  \includegraphics[height=3.5in,width=4in]{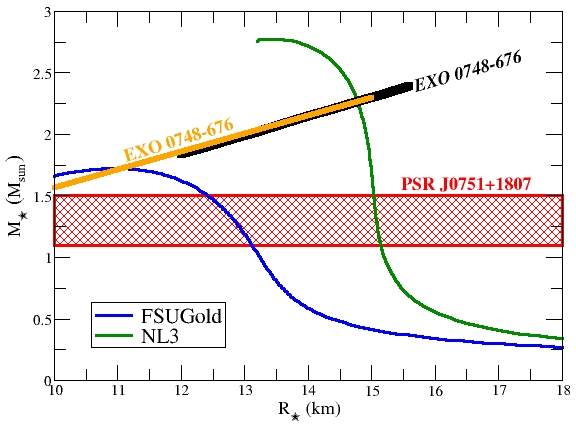}
 \vspace{-0.2cm}
\caption{(color online) Constraints on the mass-vs-radius
         relationship of neutron stars. Displayed in red is
         the {\sl recently revised} region allowed by the 
         analysis of Nice and collaborators~\cite{Nice:2005fi}. 
         The black and orange solid lines result from the analyzes
         of EXO 0748-676 by \"Ozel~\cite{Ozel:2006bv}, and
         Villarreal and Strohmayer~\cite{Villarreal:2004nj},
         respectively. Also shown are the theoretical predictions
         from the NL3~\cite{Lalazissis:1996rd,Lalazissis:1999}
         (green line) and FSUGold~\cite{Todd-Rutel:2005fa}
         (blue line) models.}
\label{Fig5}
\end{figure}

A recent observation that seems to suggest a hard equation of state is
that of the low-mass X-ray binary EXO 0748-676. Note that such a
binary system consists of a neutron star accreting mass from a normal
(non-compact) companion.  The first constraint on the equation of
state from such an object came from the detection of gravitationally
redshifted absorption lines in Oxygen and Iron by Cottam and
collaborators~\cite{Cottam:2002cu}. By measuring a gravitational
redshift of $z\!=\!0.35$, the mass-to-radius {\sl ratio} of the
neutron star gets fixed at $M/R\!\simeq\!0.15$ (with $M$ expressed in
solar masses and $R$ in kilometers). By incorporating additional
constraints arising from Eddington and thermal fluxes, a recent
analysis by \"Ozel seems to place {\sl simultaneous} limits on the
mass and radius of the neutron star in EXO 0748-676. That is,
$M\!\ge\!2.10\!\pm\!0.28~M_{\odot}$ and
$R\!\ge\!13.8\!\pm\!1.80$~km~\cite{Ozel:2006bv}. These limits are
indicated by the black solid line in Fig.~\ref{Fig5}. An earlier
determination of the spin frequency of the same neutron star by
Villarreal and Strohmayer~\cite{Villarreal:2004nj}, when combined with
the rotational broadening of surface spectral lines, yields an
independent determination of the stellar radius of
$R\!\approx\!11.5^{+3.5}_{-2.5}$~km. This estimate, when combined with
the gravitational redshift, yields the orange line in
Fig.~\ref{Fig5}. Finally, {\sl mass-vs-radius} predictions from the
NL3 and FSUGold models are displayed in Fig.~\ref{Fig5}.  The results
clearly indicate the significantly harder character of the equation of
state predicted by NL3 relative to FSUGold. This, even when both
models predict practically identical properties for existent
ground-state observables of finite nuclei.

A critical observation that would have impacted significantly on the
high-density component of equation of state is the one by Nice and
collaborators at the Arecibo radio telescope~\cite{Nice:2005fi}. Such
(initial) observation of a neutron-star--white-dwarf binary system
suggested a neutron-star mass of $M({\rm
PSR~J0751\!+\!1807})\!=\!2.1\!\pm\!0.2~M_{\odot}$. This was the
largest neutron-star mass ever reported and promised, provided that
the errors could be tighten further, to practically pin down the
high-density component of the equation of state. However, at a very
recent conference celebrating the 40th anniversary of the discovery of
pulsars in Montreal, Nice reported a significantly reduced value for
the mass of PSR~J0751\!+\!1807, namely, $M({\rm
PSR~J0751\!+\!1807})\!\approx\!1.3\!\pm\!0.2~M_{\odot}$. This revised
result is denoted by the red hatched region in Fig.~\ref{Fig5} and no
longer invalidates any of the theoretical models under consideration.

\section{Conclusions}
\label{Conclusions}

Neutron stars are unique laboratory for the study of cold baryonic
matter over an enormous range of densities. After an introduction to
the ``anatomy'' of a neutron star, I relied on recent theoretical,
experimental, and observational constraints to elucidate important
features of the equation of state of neutron-rich matter. As mentioned
in the Introduction, the only physics that spherically-symmetric
neutron stars in hydrostatic equilibrium are sensitive to is the
equation of state of neutron-rich matter [see Eqs.~(\ref{TOV})].  This
makes neutron stars gold mines for the study of baryonic matter.  The
various constraints utilized in this contribution emerged from the
universal behavior of dilute Fermi gases with large scattering
lengths~\cite{Schwenk:2005ka}, heavy-ion experiments that probe the
high-density domain of the equation of
state~\cite{Danielewicz:2002pu}, and astronomical observations that
place limits on masses and radii of neutron stars~\cite{Nice:2005fi,
Ozel:2006bv}. On the basis of these comparisons, it was concluded that
FSUGold meets all the challenges, even when no attempt was ever made
to incorporate these constraints into the calibration procedure. The
promise of new terrestrial laboratories (such as {\sl Facilities for
Rare Ion Beams}) together with improved observations with existent and
future missions (such as {\sl Constellation X}) offers the greatest
hope for determining the equation of state of cold baryonic matter in
the near future.


\begin{acknowledgments}
 The author is grateful to the organizers of the {\sl XXXI Symposium
 on Nuclear Physics} for their kind invitation and hospitality. The 
 author also wishes to acknowledge his many collaborators that were 
 involved in this work --- especially Prof. C.J. Horowitz. This work 
 was supported in part by United States Department of Energy under 
 grant DE-FD05-92ER40750.
\end{acknowledgments}

\bibliography{ReferencesJP}

\end{document}